%% file: main.tex
\documentclass[letter,reqno]{qt}

\title{
Ergodic Theorems for Quantum Trajectories
    \\
under Disordered Generalized Measurements
    }
\usepackage{authblk}
\usepackage{orcidlink}

\author[1]{Owen Ekblad\orcidlink{0009-0006-0834-0327}\thanks{ekbladow@msu.edu}}
\author[2]{Eloy Moreno-Nadales\orcidlink{0009-0001-8178-7786}\thanks{morenon4@msu.edu}}
\author[3]{Lubashan Pathirana\orcidlink{0009-0000-2130-700X}\thanks{lpk@math.ku.dk}}
\affil[1,2]{Department of Mathematics, Michigan State University, USA.}
\affil[3]{Department of Mathematical Sciences and QMATH, University of Copenhagen, Denmark.}
\date{}

\begin{document}
    \pagenumbering{arabic}
    \lhead{\thepage}
    
    \maketitle

    \begin{abstract}
    We consider quantum trajectories arising from disordered, repeated generalized measurements, which have the structure of Markov chains in random environments (MCRE) with dynamically-defined transition probabilities; we call these \emph{disordered quantum trajectories}.
    Under the assumption that the underlying disordered open quantum dynamical system approaches a unique equilibrium in time averages, we establish a strong law of large numbers for measurement outcomes arising from disordered quantum trajectories, which follows after we establish general annealed ergodic theorems for the corresponding MCRE. 
    The type of disorder our model allows includes the random settings where the disorder is i.i.d. or Markovian, the periodic (resp. quasiperiodic) settings where the disorder has periodic (resp. quasiperiodic) structure, 
    and the nonrandom setting where the disorder is constant through time. 
    In particular, our work extends the earlier noise-free results of K\"ummerer and Maassen to the present disordered framework. 
    \end{abstract}

\setcounter{tocdepth}{4}
\setcounter{secnumdepth}{4}
%\tableofcontents

\input{1_Introduction}

\input{2_Preliminaries}

\input{3_Proofs}

\section*{Acknowledgments}

    OE and EM-N completed this work while partially supported by research assistantships under Professor Jeffrey Schenker with funding from the US National Science Foundation under Grant No. DMS-2153946.

% \addcontentsline{toc}{section}{References}
\printbibliography
\end{document}

%% file: 1_Introduction.tex
\section{Introduction}
    An open quantum system, as opposed to an isolated system, is a quantum system that interacts with an external environment. 
    Given an open quantum system described by the quantum state $\rho_s$ at time $s$, the dynamics is described by a family of \emph{completely positive, trace preserving} (CPTP) super-operators $(\phi_{t,s})_{s \leq t}$ by the rule
    \begin{equation}
        \text{$\rho_t = \phi_{t,s}(\rho_s)$ given $s \leq t$.}
    \end{equation}
    The maps $\{\phi_{t,s} : s<t\}$ are referred to as \emph{dynamic propagators} or \emph{dynamical maps}. 
    If the dynamic propagators satisfy the composition law
    \begin{equation}\label{eq:cp_inhomo}
                \phi_{t,r} = \phi_{t,s}\phi_{s,r} \quad \forall \, r \leq s \leq t,
    \end{equation}
    The open quantum system is said to be \emph{Markovian}, or more specifically, \emph{time-inhomogeneous Markovian}.
    It is well-known that a family of dynamic propagators satisfying the composition law \Cref{eq:cp_inhomo} is generated by a time-dependent Lindbladian $\mcL$---also known as a GKLS generator \cite{Lindblad_1976, Gorini_1976}---such that the dynamics of the system is governed by the \emph{quantum master equation}
        \begin{equation}\label{eq_time_dep_master}
            \diff{\rho(t)}{t} 
                = \mcL(t)\rho(t), \quad \rho(t_0) = \rho_0 \, ,
        \end{equation}
    and one may obtain the dynamic propagators via
        \begin{equation}\label{eq:solution_for_master}
            \phi_{t,s} 
                = \mcT\left\{
                    \text{exp}\left(\int_{s}^t \mcL(r) \ dr\right)
                    \right\}\, ,
        \end{equation}
    where $\mcT$ denotes the time-ordering. 
    We refer the reader to \cite{rivas2012open, breuer2002theory} for more details. 
    It is possible to derive or approximate this type of master equation, and therefore the dynamics of the open system, from an intrinsically discrete dynamic model \cite{Ziman_2005, Attal_2006, Pellegrini_2010, Pellegrini_2008, https://doi.org/10.48550/arxiv.0709.3713} as sequences of repeated interactions or sequences of collisions between the quantum system and the environment. 
    Such discrete dynamic models are referred to as \textit{repeated (or sequential) interactions models} \cite{Attal_2006, Attal_2007, Strasberg_2017, Bruneau_2014}, 
    \textit{quantum collision models} \cite{Ciccarello_2022, Cattaneo_2022, Campbell_2021}, 
    or \textit{repeated (or sequential) indirect measurement models} \cite{NECHITA_2009}. 
   
        In the repeated or sequential measurement model, one assumes that the dynamics of the open system are given by a \textit{discrete} family $\{\phi_{n,m} : m\leq n \in \mbN_0\}$ of dynamic propagators. 
        If, moreover, the system is Markovian, these dynamic propagators must satisfy the composition law \Cref{eq:cp_inhomo}, so, letting $\phi_n$ denote $\phi_{n, n-1}$, we find that 
        \begin{equation}
            \phi_{n, m} 
                =
            \phi_{n}\circ\cdots\circ \phi_{m+1} \quad\forall m \leq n\in\mbN_0,
        \end{equation}
        i.e., the discrete time dynamics are determined by the one-parameter family $\{\phi_i : i\in\mbN\}$. 
        Since each $\phi_{n,m}$ is a CPTP super-operator (commonly referred to as a \textit{quantum channel}) each $\phi_i$ is also a CPTP super-operator. 
        In this work, we consider the finite-dimensional setting, where our open quantum system is described by a finite-dimensional Hilbert space.
        It is well known that, in this finite-dimensional setting, CPTP super-operators admit a Kraus form representation in terms of a collection $\{\KrausOp{i}{a}\}_{a\in\mcA_i} \subseteq \matrices$ of $d\times d$ matrices with $\mbC$-valued entries, called \textit{Kraus operators}, such that
            \begin{equation}
                \phi_i(\vblcdot)
                    :=
                    \sum_{a\in\mcA_i}
                        \KrausOp{i}{a} (\vblcdot)\KrausOp{i}{a}\adj\, \quad \text{with}\quad  \sum_{a\in\mcA_i} \KrausOp{i}{a}\adj \KrausOp{i}{a} = \mbI\, ,
            \end{equation}
        where $\mcA_i$ is some indexing set with $|\mcA_{i}|\leq d^2$ and $\mbI\in\matrices$ is the identity matrix \cite{Choi1975CompletelyMatrices}.
        The Kraus operators admit the following interpretation:
        Given the system is described by state $\rho_{i-1}$ at time $i-1$, a measurement $\phi_i$ of the system transforms to the state $\rho_i$, which, in accordance with the Born rule \cite{born1926}, is given by the probabilistic rule
            \begin{equation}
            \rho_i 
                = 
            \dfrac
                {
                \KrausOp{i}{a} \rho_{i-1} \KrausOp{i}{a}\adj
                }
                {
                \tr{\KrausOp{i}{a} \rho_{i-1} \KrausOp{i}{a}\adj}
                } 
            \quad 
                \text{ with probability } 
            \quad 
                \tr{\KrausOp{i}{a} \rho_{i-1} \KrausOp{i}{a}\adj}\, .
            \end{equation}
        The quantity $\tr{\KrausOp{i}{a} \rho_{i-1} \KrausOp{i}{a}\adj}$ is the probability of observing the outcome $a\in\mcA_i$ under the measurement $\phi_i$ in state $\rho_{i-1}$.
        In particular, if the state of the system is $\rho$ at time $0$, after $n$-many measurements, the state $\rho_n$ of the system is probabilistically described by
        \begin{equation}\label{Eqn:Intro:Markov_Chain}
            \rho_{n; \bar{a}}
                = 
            \dfrac
                {
                \KrausOpntimes{\bar{a}} \rho \KrausOpntimes{\bar{a}}\adj
                }
                {
                \tr{\KrausOpntimes{\bar{a}} \rho \KrausOpntimes{\bar{a}}\adj}
                } 
            \quad 
                \text{ with probability } 
            \quad 
                \tr{\KrausOpntimes{\bar{a}} \rho \KrausOpntimes{\bar{a}}\adj}\, ,
            \end{equation}
        where $\bar{a} = (a_1, \dots, a_n)\in \mcA_1\times\cdots\times\mcA_n$ and $\KrausOpntimes{\bar{a}} = \KrausOp{n}{a_n}\cdots \KrausOp{1}{a_1}$.
        The quantity $\tr{\KrausOpntimes{\bar{a}} \rho \KrausOpntimes{\bar{a}}\adj}$ is interpreted as the probability of observing the sequence $(a_1,\ldots a_n)$ of measurement outcomes upon conducting the sequence $(\phi_1, \dots, \phi_n)$ of measurement operations, given our system is initially described by the quantum state $\rho$. 
        Using the probabilities $\tr{\KrausOpntimes{\bar{a}} \rho \KrausOpntimes{\bar{a}}\adj}$ on finite sequences of measurement outcomes, one recovers a $\rho$-dependent probability measure $\mbQ_\rho$ on $\prod_{i\in\mbN}\mcA_i$. 
        In this way, a quantum state-valued Markov chain is defined on the probability space $\left(\prod_{i\in\mbN}\mcA_i, \otimes_{i\in\mbN}\Sigma_i, \mbQ_\rho\right)$, where $\Sigma_i$ is the discrete $\sigma$-algebra on $\mcA_i$, $\otimes_{i\in\mbN}\Sigma_i$ is the corresponding product $\sigma$-algebra, and, given initial state $\rho$, the Markov chain is given by the sequence $(\rho_n)_{n\in\mbN}$ defined in \Cref{Eqn:Intro:Markov_Chain}.
        We call this Markov chain the \textit{quantum trajectory} defined by $\{\phi_i\,\,:\,\,i\in\mbN\}$, given initial state $\rho$.
        In this article, we consider quantum trajectories obtained from a repeated measurement process under classical time-dependent disorder defined by a stationary, ergodic dynamical system. 
        That is, we assume the quantum trajectories are obtained by repeating the same measurement in a disordered environment, where the time-dependent disorder is modeled by a probability space $\seq{\Omega, \mcF, \pr}$  whose dynamics are given by a $\pr$-preserving, ergodic transformation $\theta:\Omega\to\Omega$. 
        Thus, we consider a finite collection of \textit{random} Kraus operators $\omega\mapsto \mcV_\omega=\{v_{a; \omega}\}_{a\in\mcA}$ for a fixed finite set $\mcA$. 
        We study the quantum trajectories obtained by the random sequential measurements $(\phi_{n;\omega})_{n\in\mbN}$, where 
            \begin{equation}\label{eq:measurment_process}
                \phi_{n;\omega}(\vblcdot) = \sum_{a\in \mcA} v_{a;\theta^n(\omega)} (\vblcdot) v^\dagger_{a;\theta^n(\omega)}\, .
            \end{equation}
        The sequence $(\phi_{n;\omega})_{n\in\mbN}$ is referred to as the \emph{disordered measurement process} for the collection $\mcV$, in disorder realization $\omega\in\Omega$, and $(\phi_n)_{n\in\mbN}$ is called the \emph{disordered measurement process for $\mcV$}.
        In this setting, we see that, given a disorder realization $\omega\in\Omega$ and an initial state $\rho$, after $n$-many sequential measurements the state $\rho_{n;\omega}$ of the system is probabilistically described by 
         \begin{equation}
            \rho_{n; \bar{a}; \omega}
                = 
            \dfrac
                {
                \KrausOpntimes{\bar{a}; \omega} \rho \KrausOpntimes{\bar{a}; \omega}\adj
                }
                {
                \tr{\KrausOpntimes{\bar{a}; \omega} \rho \KrausOpntimes{\bar{a}; \omega}\adj}
                } 
            \quad 
                \text{ with probability } 
            \quad 
                \tr{\KrausOpntimes{\bar{a}; \omega} \rho \KrausOpntimes{\bar{a}; \omega}\adj}\, ,
            \end{equation}
            where $\bar{a}\in\mcA^n$ and
            \begin{equation}
                \KrausOpntimes{\bar{a}; \omega}
                =
                v_{a_n; \theta^{n}(\omega)}\cdots v_{a_1; \theta(\omega)},
            \end{equation}
            and, therefore, $\tr{\KrausOpntimes{\bar{a}; \omega} \rho \KrausOpntimes{\bar{a}; \omega}\adj}$ is interpreted as the probability that we observe the first $n$ sequence of measurement outcomes $(a_1, \ldots a_n)$ under the sequence $(\phi_{1;\omega}, \ldots \phi_{n;\omega})$ of measurements, given that the disordered environment $\Omega$ is configured according to $\omega$ and the quantum system is initially in state $\rho$. 
            Similarly to before, the quantities $\tr{\KrausOpntimes{\bar{a}; \omega} \rho \KrausOpntimes{\bar{a}; \omega}\adj}$ define an $\omega$- and $\rho$-dependent probability measure on $\seq{\mcA^\mbN, \Sigma}$, where for any $n\in\mbN$ and $a_1, \dots, a_n\in\mcA$, we have that 
            \begin{equation}
                \mbQ_{\rho; \omega}\big(
                    \{\bar{a}\in\mcA^{\mbN}\,\,:\,\,
                        \pi_n(\bar{a}) = (a_1, \dots, a_n)
                    \}
                \big)
                =
                \tr{\KrausOpntimes{\bar{a}; \omega} \rho \KrausOpntimes{\bar{a}; \omega}\adj};
            \end{equation}
            here, $\Sigma$ denotes the $\sigma$-algebra generated by finite cylinder sets. 
            Of course, the construction is unchanged if one replaces $\rho$ with an $\omega$-dependent random quantum state $\omega\mapsto\rho_\omega$.
            In the following, we denote the set of all random quantum states by $\rstates$. 
            Under this construction, the disordered quantum trajectory has the structure of a Markov chain defined on the $\left(\outcomes, \Sigma \right)$ with $\omega$-dependent transition kernels;  where at each measurement the state at time $n-1$, $\rho_{n-1;\omega}$, transition to a state $\rho_{n; \omega}$ given by
            \begin{equation}\label{eq:transitions}
                 \rho_{n;\omega} 
                     = \dfrac{%   
                         v_{a_n;\theta^n(\omega)} 
                             \rho_{n-1;\omega} 
                         v\adj_{a_n;\theta^n(\omega)}
                         }%
                         {%
                         \tr{%
                             v_{a_n;\theta^n(\omega)} 
                                 \rho_{n-1;\omega} 
                             v\adj_{a_n;\theta^n(\omega)}}
                         } 
                     \quad 
                         \text{ with probability } \quad \tr{v_{a_n;\theta^n(\omega)} \rho_{n-1} v\adj_{a_n;\theta^n(\omega)}}\, .
            \end{equation}
            such Markov chains with random transition kernels are called \textit{Markov chains in a random environment} \cite{Cogburn_1980,Orey_1991}. 
            For such Markov chains, the extended random probability measure $\mbQ_{\rho;\omega}$ is called the quenched probability measure for the Markov chain started at $\rho$. 
            For ergodic type results, one usually consider the skew-product Markov chain defined on $\Omega\times\mcA^\mbN$ given by
            \begin{equation}
                \begin{split}
                    \overline{\rho_n}:\Omega\times\mcA^\mbN &\to \mbS_d\\
                    (\omega, \bar{a}) &\mapsto \rho_{n; \omega, \bar{a}}.
                \end{split}
            \end{equation}
            together with the skew-product transformation $\tau$ and the \emph{annealed quantum probability} measure. Here,
                \begin{equation}
                 \begin{split}
                \tau:\Omega\times\outcomes &\to \Omega\times\outcomes\\
                (\omega,\bar{a}) &\mapsto (\theta(\omega), \sigma(\bar{a}))
                \end{split}
            \end{equation}
            where $\sigma$ is the left shift on sequences in $\outcomes$, and the \emph{annealed quantum probability} measure for a $\Gamma\in\mcF\otimes\Sigma$ defined by
            \begin{equation}
                \overline{\mbQ_\rho}(\Gamma)
                    :=
                \int_{\Omega} 
                \mbQ_{\rho; \omega}(\Gamma^\omega)\,\dee\pr(\omega),
            \end{equation}
            where $\Gamma^\omega$ denotes the $\omega$-sections of $\Gamma$, i.e., $\Gamma^\omega := \{\bar{a}\in\mcA^\mbN\,\,:\,\, (\omega, \bar{a})\in\Gamma\}$. 
\subsection{Main results}
We assume that $\seq{\Omega, \mcF, \pr, \theta}$ is an invertible, ergodic $\pr$-preserving dynamical system, meaning that $\theta:\Omega\to\Omega$ is a $\pr$-preserving measurable bijection satisfying that
\begin{equation}
    \text{for all $F\in\mcF$, $\theta^{-1}(F) = F$ implies $\pr(F)\in\{0, 1\}$.}
\end{equation}
Let $\mcV:\Omega\to\{\text{Kraus operators}\}$ be a finite collection of random Kraus operators, which we write as
\begin{equation}
    \mcV_\omega = \{v_{a; \omega}\}_{a\in\mcA},
\end{equation}
where $\mcA$ is a fixed, finite indexing set, as above. 
We call such $\mcV$ a \textit{random Kraus ensemble} indexed by $\mcA$. 
We assume that the disordered measurement process $\seq{\phi_{i;\omega}}_{i\in\mbN}$ corresponding to $\mcV$ in \Cref{eq:measurment_process} satisfies the following condition:
\begin{description}
\hypertarget{Dyn_Erg}{}
\item[\Dynerg] 
There is a unique $\SteadyStateNoOmega\in\rstates$ such that $\phi_{1; \omega}(\SteadyState{\omega}) = \SteadyState{\theta(\omega)}$ $\pr$-almost everywhere.
\end{description}
We say $\mcV$ satisfying \Dynerg defines a \emph{dynamically ergodic} disordered quantum measurement process, and we say that $\SteadyStateNoOmega$ as above is the \emph{unique stochastically stationary state} for $\mcV$.
Our main result is establishing the following law of large numbers for disordered quantum measurement processes. 
For $n, m\in\mbN$ and $b_1, \dots, b_m\in\mcA$, let $A_n(b_1, \dots, b_m)$ denote the subset of $\mcA^\mbN$ defined by
\begin{equation}
    A_n(b_1, \dots, b_m)
        :=
    \left\{
        \seq{a_i}_{i\in\mbN}\in\mcA^\mbN
            \,\,:\,\,
        a_n = b_1, \dots, a_{n + m -1} = b_m
    \right\}.
\end{equation}
That is, $A_n(b_1, \dots, b_m)$ is the set of those sequences $\bar{a}\in\mcA^\mbN$ of measurement outcomes in which the sequence $(b_1, \dots, b_m)$ of measurement outcomes occurs starting at time $n$. 
Let $ \avg{\pr}{\mbQ_{\SteadyStateNoOmega}}(b_1, \dots, b_m)$ denote the quantity 
\begin{equation}
     \avg{\pr}{\mbQ_{\SteadyStateNoOmega}}(b_1, \dots, b_m)
        :=
    \int_\Omega 
    \mbQ_{\SteadyStateNoOmega; \omega}\big(\{
        \seq{a_i}_{i\in\mbN}\in\mcA^\mbN
            \,\,:\,\, 
        (a_1, \dots, a_m) = (b_1, \dots, b_m)
      \}\big)
      \,
      \pr(\dee\omega).
\end{equation}
Then we have the following.
\begin{restatable}[Law of Large Numbers for Measurement Outcomes of Disordered Quantum Trajectories]{thm}{thmcoutinglago}\label{Thm:Path averages equal quantum averages of equilibrium state}
Assume that $\seq{\Omega, \mcF, \pr, \theta}$ is an invertible, ergodic $\pr$-preserving dynamical system, and let $\omega\mapsto \mcV_\omega$ define a random Kraus ensemble indexed by $\mcA$. 
If $\mcV$ satisfies \Dynerg with unique stochastically stationary state $\SteadyStateNoOmega$, then for any $\vartheta\in\rstates$, $\pr$-almost every $\omega\in\Omega$, and every $b_1, \dots, b_m\in\mcA$, we have that
            \begin{equation}\label{slln:eq}
                \lim_{N\to\infty}
                \cfrac{\#\Big\{
                n < N\,\,:\,\, 
                \bar{a}\in A_n(b_1, \dots, b_m)
            \Big\}}{N}
                    =
                \avg{\pr}{\mbQ_{\SteadyStateNoOmega}}(b_1, \dots, b_m)
            \end{equation}
            holds for $\mbQ_{\vartheta; \omega}$-almost every $\bar{a}\in\outcomes$.
\end{restatable}
So, with total freedom in how the quantum system and the disordered environment is configured, by repeatedly measuring our system according to $\mcV$ and recording the outcome, we may compute $\avg{\pr}{\mbQ_{\SteadyStateNoOmega}}(b_1, \dots, b_m)$ for any $b_1, \dots, b_m\in\mcA$, which, in turn, enables us to compute $\avg{\pr}{\mbQ_{\SteadyStateNoOmega}}$.
To prove this theorem, we establish more general ergodic theorems, which we now describe. 
Recall that $\Sigma$ is the $\sigma$-algebra on $\mcA^\mbN$ generated by all finite cylinder sets (see \cref{sec:prelim}). 
Let $\sigma:\mcA^\mbN\to\mcA^\mbN$ be the left shift map $\bar{a} = \seq{a_i}_{i\in\mbN}\mapsto \seq{a_{i+1}}_{i\in\mbN}$ and let $\tau:\Omega\times\mcA^\mbN\to \Omega\times\mcA^\mbN$ be the skew-shift map $\tau(\omega, \bar{a}) = (\theta(\omega), \sigma(\bar{a}))$. 
We then have the following.
\begin{restatable}[Annealed Ergodic Theorem]{thm}{thmAnnnealedErgodicity}\label{Thm:Annealed_Erg_Thm_1}
         Assume that $\seq{\Omega, \mcF, \pr, \theta}$ is an invertible, ergodic $\pr$-preserving dynamical system, and let $\omega\mapsto \mcV_\omega$ define a random Kraus ensemble indexed by $\mcA$. 
         If $\mcV$ satisfies \Dynerg with unique stochastically stationary state $\SteadyStateNoOmega$, then 
         \begin{enumerate}[label=(\roman*)]
             \item $\tau$ is a $\bQ_{\SteadyStateNoOmega}$-preserving ergodic transformation.  
             That is, for any $\Gamma\in\mcF\otimes\Sigma$, we have that $\bQ_{\SteadyStateNoOmega}(\tau^{-1}(\Gamma)) = \bQ_{\SteadyStateNoOmega}(\Gamma)$, and, moreover, $\tau^{-1}(\Gamma) = \Gamma$ implies that 
             \begin{equation}
                 \bQ_{\SteadyStateNoOmega}(\Gamma)\in\{0, 1\}. 
             \end{equation}
             \item Furthermore, if $\tau^{-1}(\Gamma) = \Gamma$, then for any $\vartheta\in\rstates$, we have that
             \begin{equation} 
             \bQ_{\vartheta}(\Gamma)
                =
             \bQ_{\SteadyStateNoOmega}(\Gamma)\in\{0, 1\}. 
             \end{equation}
             That is, the quantity $\bQ_{\vartheta}(\Gamma)$ is independent of $\vartheta$, and is equal to either 0 or 1.
         \end{enumerate} 
\end{restatable}
The primary content of this theorem is the ergodicity of $\bQ_{\SteadyStateNoOmega}$ for $\tau$.
Along the way to proving this theorem, we establish the following result of independent interest. 
\begin{restatable}[Law of Large Numbers for Annealed Quantum Probability]{thm}{thmlln}\label{Thm:lln_for_Q}
Assume that $\seq{\Omega, \mcF, \pr, \theta}$ is an invertible, ergodic $\pr$-preserving dynamical system, and let $\omega\mapsto \mcV_\omega$ define a random Kraus ensemble indexed by $\mcA$. 
If $\mcV$  satisfies \Dynerg with unique stochastically stationary state $\SteadyStateNoOmega$, then for any $\vartheta\in\rstates$, we have that 
\begin{equation}
    \lim_{N\to\infty}\dfrac{1}{N}\sum_{n=1}^N \bQ_{\vartheta} (\tau^{-n}(\Gamma)) = \bQ_{\SteadyStateNoOmega}(\Gamma)\,
\end{equation}
for all $\Gamma\in\mcF\otimes\Sigma$. 
\end{restatable}
By considering sets $\Gamma\in\mcF\otimes\Sigma$ of the form $\Gamma = \Omega\times E$ where $E\in\Sigma$ is $\sigma$-invariant, we recover the following quenched ergodic theorem. For $\vartheta\in\rstates$, let $\avg{\pr}{\mbQ_{\SteadyStateNoOmega}}$ denote the measure on $\Sigma$ defined by
\begin{equation}
    \avg{\pr}{\mbQ_{\vartheta}}(E) := \int_\Omega \mbQ_{\vartheta; \omega}(E) \dee\pr(\omega)
\end{equation}
for $E \in \Sigma$.
\begin{restatable}[Quenched Ergodic Theorem]{thm}{thmQuenchedErg}\label{Thm:Quenched_Ergodic}
Assume that $\seq{\Omega, \mcF, \pr, \theta}$ is an invertible, ergodic $\pr$-preserving dynamical system, and let $\omega\mapsto \mcV_\omega$ define a random Kraus ensemble indexed by $\mcA$. 
If $\mcV$ is satisfies \Dynerg with unique stochastically stationary state $\SteadyStateNoOmega$, then 
\begin{enumerate}[label=(\roman*)]
    \item $\sigma$ is a $\avg{\pr}{\mbQ_{\SteadyStateNoOmega}}$-preserving ergodic transformation. 
    That is, for any $E\in\Sigma$, we have that $\avg{\pr}{\mbQ_{\SteadyStateNoOmega}}\seq{\sigma^{-1}\seq{E}}
    =
    \avg{\pr}{\mbQ_{\SteadyStateNoOmega}}\seq{E}$, and, moreover, $\sigma^{-1}(E) = E$ implies that 
    \begin{equation}
        \avg{\pr}{\mbQ_{\SteadyStateNoOmega}}\seq{E}\in\{0, 1\}.
    \end{equation}
    \item Furthermore, if $\sigma^{-1}(E) = E$, then for any $\vartheta\in\rstates$, we have that 
    \begin{equation}
    \mbQ_{\vartheta; \omega}\seq{E}
        =
    \mbE_{\pr}[\mbQ_{\SteadyStateNoOmega}]\seq{E}
    \in\{0, 1\}.
    \end{equation}
    That is, the quantity $\mbQ_{\vartheta; \omega}\seq{E}$ is $\pr$-almost surely constant, and this constant is independent of $\vartheta$ and is equal to either 0 or 1. 
\end{enumerate}
\end{restatable}
\subsection{Related literature}
This work may be viewed as an extension of the discrete time results in \cite{Kummerer1999AnMeasurement, kummerer2003ergodic} to the setting with the above-described type of disorder present.
Often, quantum trajectories are modeled by a \textit{continuous}-time stochastic process satisfying the stochastic Schr\"odinger equations: the interested reader may consult any of the sources \cite{Davies1976QuantumSystems, Gisin1984QuantumProcesses, Carmichael1993An1991,  Holevo2001StatisticalTheory, Bouten2004StochasticEquations, Barchielli2009QuantumCase, Wiseman2009QuantumControl} for further reading on the general theory of quantum trajectories. 
As discussed above, the discrete-time stochastic processes of the sort we are considering provide a good approximation of these continuous-time processes, in a sense made precise in \cite{Pellegrini_2008, Bauer2012RepeatedLimit}.
The formalism of quantum trajectories models various experiments in quantum optics \cite{Carmichael1993An1991, Haroche2006ExploringQuantum, Wiseman2009QuantumControl}---notably, the Nobel prize winning work of Serge Haroche's group \cite{Guerlin2007ProgressiveCounting} is modeled by quantum trajectories. 
Various authors have studied the limiting behavior and attractors of quantum trajectories, achieving various sorts of ergodic theorems and laws of large numbers. 
For more reading on this area, the reader may consult the work \cite{Benoist2023LimitTrajectories}, the introduction of which contains a succinct survey of the area. 
The field of disordered open quantum dynamical systems is actively developing, and the interested reader may consult the non-exhaustive list of related works 
\cite{Bruneau2008RandomSystems, Pellegrini2009Non-MarkovianMeasurements, Bruzda2009RandomOperations, Bruneau2010InfiniteDynamics, Nechita2012RandomStates, Burgarth2013ErgodicDimensions, Bruneau_2014, Movassagh2021TheoryProcesses, Bougron2022MarkovianSystems, Movassagh2022AnStates, Pathirana2023LawProcesses, Schenker2024QuenchedMeasurements, ekblad2024ergodic, ekblad2024asymptotic}.
In \cite{ekblad2024asymptotic}, particularly, the specific model of disordered quantum trajectories we are considering was introduced. 
%
%
%
% As described above, the disordered quantum trajectory is an instance of a time-inhomogeneous Markov chain in a random environment, and one may consult \cite{Cogburn_1980, Orey_1991} for further reading on this topic. 
%%
%%

%% file: 2_Preliminaries.tex
\section{Notations and preliminary results} \label{sec:prelim}
    For a topological space $\mcX$, we denote by $\mcB(\mcX)$ the Borel $\sigma$-algebra of subsets of $\mcX$. 
    Given two measure spaces $(\mcX, \mcF)$ and $(\mcY, \mcG)$ where $\mcF$ and $\mcG$ are the two corresponding $\sigma$-algebras, we denote by $\mcF\otimes\mcG$ the product $\sigma$-algebra generated on the space $(\mcX\times \mcY)$. 
    Often $(\Omega,\mcF,\pr)$ will be used to denote an arbitrary probability space, where $\mcF$ is a $\sigma$-algebra of subsets of $\Omega$ and $\pr$ is the probability measure.

    We shall now introduce the preliminary mathematical objects and constructions used in this work. 
    Let $(\Omega,\mcF,\pr)$ be a probability space. 
    For a $(\Omega,\mcF)$-random variable or matrix $X$, we write $\avg{\pr}{X}$ to denote the expectation
        \begin{equation}
            \avg{\pr}{X} := 
            \int_\Omega X \dee{\pr}. 
        \end{equation}
    In the case that $X$ is a matrix-valued random variable, $\avg{\pr}{X}$ is defined by functional calculus.
    Given a measurable map $\theta:\Omega\to\Omega$ on the probability space, $(\Omega,\mcF,\pr)$, we say that $\theta$ is a $\pr$-\emph{probability-preserving transformation} (abbreviated as \textit{ppt}) if $\pr(\theta^{-1}(A)) = \pr(A)$ for all $A\in\mcF$.
    We say $A\in\mcF$ is $\theta$-invariant if $\theta^{-1}(A) = A$. 
    We say that a $\pr$-ppt $\theta$ is \emph{ergodic} for the probability measure $\pr$, or that $\pr$ is $\theta$-ergodic, if every $\theta$-invariant set $A\in\mcF$ satisfies
        \begin{equation}
        \pr(A)\in\{0, 1\}.
        \end{equation}
    Equivalently, any \emph{essentially $\theta$-invariant} event $A\in\mcF$ has $\pr$-probability $0$ or $1$, where an event $A\in\mcF$ is called essentially $\theta$-invariant, if $\pr(\theta^{-1}(A)\triangle A) = 0$  where $\triangle$ denotes the symmetric difference of sets. 
    We call a collection $(\Omega, \mcF, \pr, \theta)$ a probability-preserving (resp. ergodic) dynamical system if $\theta$ is a ppt (resp. ergodic ppt).
    In the following, we shall assume that $(\Omega, \mcF, \pr, \theta)$ is an ergodic dynamical system with the additional condition that $\theta$ is invertible. 
    It is straightforward to check that if $\theta$ is an invertible ergodic ppt with measurable inverse, then $\theta^{-1}$ is also an ergodic probability preserving transformations.  
    We shall freely use standard facts about ergodic ppt which we do not list here. 
    We refer the reader to standard textbooks on ergodic theory such as \cite{brin2002introduction, walters2000introduction} for reference. 
    Not all ergodic ppt are necessarily invertible, but any surjective ergodic transformation on a standard probability space extends to an invertible and ergodic; 
    such extensions are called \emph{natural extensions}, and we refer the reader to  \cite{rohlin1964exact, rohlin1961exact} or \cite[\S10]{cornfeld2012ergodic} for detailed explanations. 
    Let $\matrices$ denote the set of $d\times d$ matrices with complex entries. 
    We let $\mbI$ denote the identity matrix in $\matrices$. 
    Our state space will be the set $\states$ of positive semi-definite matrices with trace 1, 
        \begin{equation}
            \states 
                :=
            \left\{
                \vartheta\in\matrices 
                \,\,:\,\,
                \text{$\vartheta$ positive semi-definite and $\tr{\vartheta} = 1$}
            \right\}.
        \end{equation}
    We refer to $\states$ as the set of \emph{quantum states}, and refer to $\vartheta\in\states$ as a \emph{quantum state}. 
    Given $\vartheta\in\states$ and a matrix $M\in\matrices$, we let $M\bcdot\vartheta$ denote the quantum state defined by 
        \begin{equation}\label{Prelim:Eqn:Projective action}
             M\bcdot\vartheta 
                :=
            \begin{cases}
                \cfrac{M\vartheta M^\dagger}{\tr{M\vartheta M^\dagger}} &\text{if }\tr{M\vartheta M^\dagger}\neq 0\, ,\\
                d^{-1}\mbI &\text{else}.
            \end{cases}
        \end{equation}
    For $M\in\matrices$, we let $M^\dagger$ denote its conjugate transpose, and we let $|M| = \sqrt{M^\dagger M}$ denote the matrix defined by the functional calculus.
    We give $\matrices$ the structure of a Hilbert space by means of the Hilbert-Schmidt inner product 
        \begin{equation}
            \inner M L 
                :=
            \tr{M^\dagger L} \quad\text{for } M, L\in\matrices.
        \end{equation}
    Note that we have taken the convention that $\inner{\vblcdot}{\vblcdot}$ is conjugate-linear in the first entry. 
    We denote by  $\|\vblcdot\|$ the trace class norm (also known as the Schatten $1$-norm), i.e., $\norm{M} = \operatorname{Tr}|M|$.

    We call a linear map $\phi:\matrices\to\matrices$ a \emph{super-operator} and denote the space of such objects by $\mapspace$.
    The set $\mapspace$ is precisely the space of bounded operators on the Banach space $\matrices$, and so has a Banach space structure induced by the norm $\|\vblcdot\|$ on $\matrices$. 
    That is, for $\phi\in\mapspace$, 
        \begin{equation}
            \|\phi\|
                :=
                    \sup\{\norm{\phi(M)}: M\in \matrices, \norm{M}\leq 1\}
        \end{equation}
    defines a norm making $\mapspace$ into a Banach space. 
    Moreover, for $\phi\in\mapspace$ there is a unique superoperator $\phi^\dagger\in\mapspace$ satisfying 
        \begin{equation}
            \inner{\phi(M)}{L}
                =
                    \inner{M}{\phi^\dagger(L)}
        \end{equation}
    for all $M, L\in\matrices$. 
    We call $\phi^\dagger$ the \emph{adjoint} of $\phi$. 
    If $\phi\in\mapspace$ satisfies 
        \begin{equation}
            \tr{\phi(M)}
                =
                \tr{M}
                \quad\text{for all $M\in\matrices$}.
        \end{equation}
    We call $\phi$ \emph{trace-preserving}. 
    If $\phi(\rho)\geq 0$ for all $\rho\in\states$, we call $\phi$ positive; 
    if, moreover, for all $n\in\mbN$, the super-operator 
        \begin{equation}
            \phi\otimes\operatorname{Id}_n:\matrices\otimes\mbM_n\to\matrices\otimes\mbM_n
        \end{equation}
    is positive, we call $\phi$ \emph{completely positive}. 
    We call $\phi\in\mapspace$ a \emph{quantum channel} if $\phi$ is completely positive and trace preserving (henceforth abbreviated CPTP.).
    It is a standard fact (known as the Choi-Kraus theorem) that any quantum channel $\phi$ may be written in its Kraus form 
        \begin{equation}
            \phi(\vblcdot)
                :=
                    \sum_{a\in\mcA}
                    v_a(\vblcdot)v_a^\dagger
        \end{equation}
    for some finite set $\{v_a\}_{a\in\mcA}$ of matrices $V_a\in\matrices$ satisfying the relation 
    \begin{equation}
        \sum_{a\in\mcA}v_a^\dagger v_a
            =
            \mbI;
    \end{equation}
    the set $\{v_a\}_{a\in\mcA}$ is called the collection of \emph{Kraus operators} for the quantum channel $\phi$. 
    It is not hard to see then that the adjoint $\phi^\dagger$ of a quantum channel $\phi$ is defined by 
        \begin{equation}
             \phi^\dagger(\vblcdot)
                =
                    \sum_{a\in\mcA}
                        v_a^\dagger(\vblcdot)v_a. 
        \end{equation}
    We refer the reader to \cite{watrous2018theory} for more details on quantum channels and related matters. 
    In the following, we shall frequently refer to various sets of random matrices. 
    We let $\rmatrices$ denote the set of random matrices, and let $\rstates$ denote the subset of $\rmatrices$ consisting of those matrices that are almost surely elements of $\states$. 
    For $1\leq p\leq \infty$, we write $L^p(\Omega,\mbM_d)$ to denote the subset of $\rmatrices$ that has $L^p(\Omega)$ operator norms, i.e.  
        \begin{equation}
            L^p(\Omega,\mbM_d)
                := 
                    \{ 
                        X\in\rmatrices \: : \:
                        \omega\mapsto \|X(\omega)\|\in L^p(\Omega)
                    \}\, .
        \end{equation} 
    \subsection{Disordered quantum trajectories and related concepts}
    In this work, we consider certain time-inhomogenous Markov chains in a random environment with state space $\states$, which we call \textit{disordered quantum trajectories}. 
    We now devote some time to rigorously defining disordered quantum trajectories. 
        Let $\mcA$ be a fixed finite set, which we refer to as the set of possible outcomes, and let $\mcV = \{v_a\}_{a\in\mcA}$ be a collection of $\matrices$-valued random variables defined on $(\Omega,\mcF,\pr)$ satisfying 
            \begin{equation}\label{Eqn:Stochasticity_condition_Kraus_ops}
                \sum_{a\in\mcA}
                    v_{a;\omega}^\dagger v_{a;\omega}
                    =
                    \mbI
            \end{equation}
        for $\pr$-almost every $\omega\in\Omega$. 
        We call such $\mcV$ a \textit{random Kraus ensemble}. 
        Given a random Kraus ensemble $\mcV = \{v_a\}_{a\in\mcA}$, we let $\phi_\mcV:\Omega\to\mapspace$ be the random quantum channel defined by
            \begin{equation}
                \phi_{\mcV; \omega}(\vblcdot)
                    := 
                        \sum_{a\in\mcA}
                        v_{a; \omega}(\vblcdot)v_{a; \omega}^\dagger.
            \end{equation}
        For shorthand, we write $T_{a}$ to denote the random super-operator defined by 
        \begin{equation}\label{eq:T-operator}
            T_{a; \omega}(M) 
                = 
                    v_{a; \omega}Mv_{a; \omega}^\dagger, 
        \end{equation}
        so that $\phi_\mcV = \sum_{a\in\mcA}T_a$ and $\phi_\mcV^\dagger = \sum_{a\in\mcA}T_a^\dagger$. 
        For the rest of this work, we fix a random Kraus ensemble $\mcV$ with corresponding quantum channel $\phi_\mcV$, and we shall simply write $\phi$ to denote $\phi_\mcV$. 
        For $n\in\mbZ$, we let $\phi_n$ be the random quantum channel defined by 
            \begin{equation}
                 \phi_{n; \omega}
                    :=
                        \phi_{\theta^n(\omega)},
            \end{equation}
        and, for any $N\in\mbZ$, we let $\Phi^{(N)}$ denote the random quantum channel defined by the composition 
            \begin{equation}
                \Phi^{(N)}
                    := 
                    \begin{cases}
                        \phi_{N} \circ\cdots\circ\phi_{1} 
                            &
                            \text{for $N\geq 1$}\\
                        \operatorname{Id}_{\matrices} 
                            &
                            \text{for $N=0$}
                            \\
                        \phi_{0}\circ\cdots\circ\phi_{N+1} 
                            &
                            \text{for $N\leq -1$}.
                    \end{cases}
            \end{equation}
        We endow the set $\outcomes$ of sequences of elements of $\mcA$ with the $\sigma$-algebra generated by the projection maps $(\pi_m)_{m\in\mbN}$ defined by
            \begin{align}
                \pi_m:\outcomes&\to\mcA^m\\
                (a_n)_{n\in\mbN}&\mapsto (a_1, \dots, a_m), 
            \end{align}
        where $\mcA^m$ is made a measure space via the discrete $\sigma$-algebra $\Sigma_m$. 
        We denote the resulting $\sigma$-algebra on $\outcomes$ by $\Sigma$.
        For $(b_1, \dots, b_m)\in\mcA^m$ and a probability measure $\mbP$ on $(\outcomes, \Sigma)$, we use the shorthand notation 
            \begin{equation}
                \mbP(b_1, \dots, b_m) 
                := 
                \mbP\big(\pi_m^{-1}\{(b_1, \dots, b_m)\}\big).
            \end{equation}
        %
        %Let $\mcF\otimes\Sigma$ be the product $\sigma$-algebra on $\Omega\times\outcomes$.
        %
        For $n\in\mbN$, we let $V_n:\Omega\times\outcomes\to\matrices$ be the $\mcF\otimes\Sigma$-measurable function defined by
            \begin{equation}\label{eq:big_V}
                V_{n; \omega}(\bar{a})
                    :=
                v_{a_n; \theta^n(\omega)}\cdots v_{a_1; \theta(\omega)} \quad\text{where $\bar{a} = (a_m)_{m\in\mbN}$.}
            \end{equation}
        By abuse of notation, we write $V_{n; \omega}(a_1, \dots, a_n)$ to denote the above expression for $a_1, \dots, a_n\in\mcA$. 
        For a random quantum state $\vartheta\in\rstates$, we let $\left(\vartheta_n\right)_{n\in\mbN}$ be the sequence of $\mcF\otimes\Sigma$-measurable functions $\Omega\times\outcomes\to\states$ defined by 
            \begin{equation}
                \vartheta_{n; \omega}(\bar{a})
                    :=
                V_{n; \omega}(\bar{a})\cdot \vartheta_\omega, 
            \end{equation}
        where $V_{n; \omega}(\bar{a})\cdot \vartheta_\omega$ was defined in \Cref{Prelim:Eqn:Projective action}.
        We refer to the sequence $\seq{\vartheta_n}_{n\in\mbN}$ of measurable functions so defined as the \emph{disordered quantum trajectory} associated to $\mcV$ with initial state $\vartheta$. 
        In this work, we are concerned with establishing ergodic properties of disordered quantum trajectories satisfying the following notion of \textit{dynamical ergodicity}. 
        \begin{dfn}[Dynamical Ergodicity]\label{dfn:stationary_state}
        Given a random quantum channel $\phi:\Omega\to\mapspace$ defined on the ergodic dynamical system $(\Omega,\mcF,\pr,\theta)$, we say that a random state $\rho\in\rstates$ is stochastically stationary under $\phi$ if 
        \begin{equation}\label{eq:statinary_sate}
                    \phi_{\theta(\omega)}(\rho_{\omega}) 
                        = 
                            \rho_{\theta(\omega)}\, \quad \pr\text{-almost surely}.
                \end{equation}
        If, up to almost sure equivalence, there is a unique $\rho\in\rstates$ satisfying \Cref{eq:statinary_sate}, we say that $\phi$ is dynamically ergodic, and we write $\SteadyStateNoOmega$ to denote its unique stochastically stationary state. 
        We say that a random Kraus ensemble $\mcV$ is dynamically ergodic with stationary state $\SteadyStateNoOmega$ if $\phi_\mcV$ is dynamically ergodic with unique stochastically stationary state $\SteadyStateNoOmega$.
        \end{dfn}

    \subsection{Associated probability measures}\label{sec:quantum_measures}
        In this section, we formally define the quantum measures $\mbQ_{\rho;\omega}$ and $\bQ_\rho$. 
        We start by noticing that for any random initial state $\rho\in\rstates$, $n\in\mbN$ and for $\pr$-almost every $\omega \in \Omega$, the quantity (for any $A \subset \mcA^n$ and $V_{n; \omega}(\bar{a})$ as in \Cref{eq:big_V})
            \begin{equation}
                \mbQ^{n}_{\rho;\omega}(A): = \sum_{\bar{a}\in A} 
            \tr{
            V_{n; \omega}(\bar{a})
                        \rho_\omega
                    V^\dagger_{n; \omega}(\bar{a})
                }
            \end{equation}
        defines a probability measure on $\mcA^n$, and, moreover,  $\left(\mbQ^{n}_{\rho;\omega}\right)_{n\in\mbN}$ is a {consistent family} of probability measures, in the sense that for any $A\subseteq\mcA^n$ we have
            \begin{equation}
                \mbQ_{\varrho;\omega}^{(n+1)}(A \times \mcA)
                    \ = \ 
                    \mbQ_{\varrho;\omega}^{(n)}(A)
                    \quad \pr\text{-almost surely},
            \end{equation}
        which is ensured by \Cref{Eqn:Stochasticity_condition_Kraus_ops}. 
        Therefore, by the Kolmogorov extension theorem \cite{kolmogoroff1933grundbegriffe}, for $\pr$-almost every $\omega\in\Omega$, there is a probability measure $\mbQ_{\rho;\omega}$ on $(\outcomes, \Sigma)$ such that for any $\bar{a}\in\mcA^n$
            \begin{equation}
                \mbQ_{\rho;\omega}\left(\pi_n^{-1}(\bar{a})\right) 
                = 
                \tr{V_{n; \omega}(\bar{a}) \rho_\omega 
                V^\dagger_{n; \omega}(\bar{a})}.
            \end{equation}
        We record this discussion as a proposition. 
        \begin{prop}\label{prop:quenched_measure}
            Given any random initial state $\rho\in\rstates$, we have that $\pr$-almost for all $\omega\in\Omega$ the existence of a probability measure $\mbQ_{\rho;\omega}$ on $(\outcomes, \Sigma)$ such that for any $n\in\mbN$ and  $A\subseteq \mcA^n$ it satisfies the formula 
            $\mbQ_{\rho;\omega}(A) = \sum_{\bar{a} \in A} \tr{V_{n; \omega}(\bar{a}) \rho_\omega V^\dagger_{n; \omega}(\bar{a})}$. 
            In particular, for $a_1, \dots, a_n\in\mcA$,
                \begin{equation}\label{eq:probaility_of_finite_string}
                    \mbQ_{\rho; \omega}\big(\pi_n^{-1}\{(a_1, \dots, a_n)\}\big)
                        =
                    \tr{V_{n; \omega}(\bar{a}) \rho_\omega 
                V^\dagger_{n; \omega}(\bar{a})}
                \end{equation} 
            holds $\pr$-almost surely.
        \end{prop}
        The same $\omega$-dependent measure was obtained in \cite{ekblad2024asymptotic} in the case that $\rho$ was nonrandom.
        We call it the quenched quantum probability measure. 
        \begin{dfn}[Quenched Quantum Probability]\label{dfn:quneched_measure}
            The probability measure $\mbQ_{\rho;\omega}$ is called the {quenched quantum measure} associated to the random initial state $\rho$ and disorder realization $\omega$.
        \end{dfn}
        There is a rephrasing of $\mbQ$ in terms of a certain random matrix-valued measure that is useful in streamlining our presentation.
        Recall that, given a measurable space $(\Xi, \mcG)$, we say a function $M:\mcG\to\matrices$ is a matrix-valued probability measure if for all quantum states $\vartheta\in\states$, the set map 
            \begin{equation}
                \mcG \ni E\mapsto \inner{\vartheta}{M(E)}
            \end{equation}
        defines a probability measure on $(\Xi, \mcG)$.
        More generally, we say a function $M:\Xi\times\mcG\to\matrices$ is a random matrix-valued probability measure if for all $\vartheta\in\states$ and for all $E\in\mcG$, the map $\Xi\ni\xi\mapsto \inner{\vartheta}{M_\xi(E)}$ is measurable and if for almost every $\xi\in\Xi$, $E\mapsto \inner{\vartheta}{M_\xi(E)}$ defines a matrix-valued probability measure.

        \begin{lemma}\label{Lem:The quantum measure}
            Let $\mcV = \{v_a\}_{a\in\mcA}$ be a random Kraus ensemble.
            There is a random matrix-valued measure $\mbQ^*$ on $\outcomes$ such that for all $n\in\mbN$ and all $a_1, \dots, a_n\in\mcA$, 
                \begin{equation}
                    \mbQ^*_\omega\big(\pi_n^{-1}\{(a_1, \dots, a_n)\}\big)
                        =
                    T_{a_1; \theta(\omega)}^\dagger\circ\cdots\circ T_{a_n; \theta^n(\omega)}^\dagger(\mbI)
                \end{equation}
            holds for almost every $\omega\in\Omega$. 
        \end{lemma}
            \begin{proof}
                Let $\rho\in\states$ be a quantum state, fix $\omega\in\Omega$, and consider the quenched quantum measure $\mbQ_{\rho;\omega}$.
                %obtained in Proposition \ref{prop:quenched_measure}. 
                %
                For each $n\in\mbN$ and all $a_1, \dots, a_n\in\mcA$, we have that 
                    \begin{align}
                        \mbQ_{\rho; \omega}\big(\pi_n^{-1}\{(a_1, \dots, a_n)\}\big)
                            &=
                        \tr{V_{n; \omega}(\bar{a}) \rho_\omega 
                                V^\dagger_{n; \omega}(\bar{a})}
                            \\
                            &=
                        \inner{\rho}{T_{a_1; \theta(\omega)}^\dagger\circ\cdots\circ T_{a_n; \theta^n(\omega)}^\dagger(\mbI)}
                    \end{align}
                holds for $\pr$-almost every $\omega\in\Omega$. 
                Thus, by defining the set map $\mbQ^*:\Sigma\to L^\infty(\Omega, \matrices)$ to be the random matrix satisfying the $\Pr$-almost sure equality
                    \begin{equation}
                        \mbQ_{\rho; \omega}(E)
                            := 
                        \inner{\rho}{\mbQ^*_\omega(E)},
                    \end{equation}
                the lemma then follows given by the structure of $\inner{\vblcdot}{\vblcdot}$ as a Hilbert space inner product on $\matrices$. 
            \end{proof}
        \begin{dfn}[Matrix-valued Quenched Quantum Probability]\label{dfn:quneched_matrix}
        We call the random matrix-valued measure $\mbQ^*$ obtained in \Cref{Lem:The quantum measure} the {matrix-valued quenched quantum measure} associated to $\mcV$. 
        \end{dfn}
        We shall freely and often use the fact that for $\pr$-almost every $\omega\in\Omega$ and every $E\in\Sigma$, the equality
        \begin{equation}
                        \mbQ_{\rho; \omega}(E)
                            := 
                        \inner{\rho}{\mbQ^*_\omega(E)},
                    \end{equation}
        holds, as this often clarifies otherwise subscript-heavy notation. 
%       %

        %        
        We now turn our attention to discussing the annealed quantum measure. 
        The disordered quantum trajectory may be viewed as a process on the joint disorder-outcome space $\Omega \times \outcomes$. 
        Informally speaking, the annealed quantum probability is a $\pr$-average of $\mbQ_{\vartheta; \omega}$ at different sections with respect to $\omega \in \Omega$. 
        Recall a technical lemma from \cite{ekblad2024asymptotic}.
        \begin{lemma}\label{lemma:quantum_prob_is_measurable}
            Given any random state $\vartheta\in\rstates$, the map 
                \begin{align}\label{Random quantum probability}
                    \begin{split}
                        \mbQ_{\vartheta}: \Omega 
                            &
                                \to \mcP\left(\outcomes\right)\\
                        \omega
                            & 
                                \mapsto \  \mbQ_{\vartheta;\omega} \ 
                    \end{split}
                \end{align}
            is a measurable map between the two measure spaces $(\Omega,\mcF)$ and $\left(\mcP\left(\outcomes\right), \mcB\left(\mcP\left(\outcomes\right)\right)\right)$. 
            Here $\mcP\left(\outcomes\right)$ denotes the set of Radon probability measures on $\outcomes$ and $\mcB\left(\mcP\left(\outcomes\right)\right)$ is the Borel $\sigma$-algebra induced by the Prokhorov metric. 
        \end{lemma}
            \begin{proof}
                This lemma was proved for a deterministic $\vartheta$ in \cite[Lemma A.2]{ekblad2024asymptotic} and the proof carries over for any random state. 
            \end{proof}

        We are now ready to define the annealed quantum probabilities. 
        
        \begin{dfn}[Annealed Quantum Probability]\label{dfn:anneaeld_measure}
            For $\Gamma\in \mcF\otimes \Sigma$ and $\omega\in\Omega$, let $\Gamma^\omega$ denote the $\omega$-section of $\Gamma$, 
            \begin{equation}
                \Gamma^\omega
                    :=
                \{\bar{a}\in\outcomes: (\omega,\bar{a}) \in \Gamma\}.
            \end{equation}  
            For $\vartheta\in\rstates$, we define the \textit{annealed quantum measure} associated to $\vartheta$ by 
            \begin{equation}\label{eq:extended_probability}
                \bQ_{\vartheta}(\Gamma) 
                    = \int_\Omega  \ 
                    \mbQ_{\vartheta; \omega}(\Gamma^\omega) \ 
                    \pr(\dee\omega).
            \end{equation}
        \end{dfn}
        In the above, the measurability of $\omega \mapsto \mbQ_{\vartheta;\omega}$ from \Cref{lemma:quantum_prob_is_measurable} is used to rigorously obtain \Cref{eq:extended_probability}.
        Recall also the map $\tau = (\theta, \sigma):\Omega\times\outcomes\to\Omega\times\outcomes$ defined by 
            \begin{equation}\label{eq:tau}
                \tau(\omega,\bar{a}) 
                    = 
                        (\theta(\omega), \sigma(\bar{a}))\, ,
            \end{equation}
        where $\sigma$ is the left-shift $\sigma\big((a_n)_{n\in\mbN}\big)
                    = 
                (a_{n+1})_{n\in\mbN}$.

%% file: 3_Proofs.tex
\section{Technical results and proofs of theorems}
We begin with some basic technical lemmas regarding the objects just introduced. 
\begin{lemma}\label{lem:disordered_k_m_lemma_1_2}
Let $S\in\Sigma$ and $n\in\mbN$. 
For $a_1, \dots, a_n\in \mcA$ and any $\vartheta\in\rstates$, we have that 
\begin{equation}\label{eq:disordered_k_m_lemma_1_2_q*}
                    \mbQ^*_\omega\seq{\sigma^{-n}(S)\cap\pi_n^{-1}\{(a_1, \dots, a_n)\}}
                        =
                    T_{a_1; \theta(\omega)}^\dagger\circ\cdots\circ T_{a_n; \theta^n(\omega)}^\dagger
                    \seq{\mbQ^*_{\theta^n(\omega)}\seq{S}}
\end{equation}
for $\pr$-almost every $\omega\in\Omega$. 
        \end{lemma}
\begin{proof}
By a monotone class argument, it suffices to prove the result for $S\in\Sigma$ of the form $S = \pi^{-1}_m\{(b_1,\ldots b_m)\}$ for any $b_1, \dots, b_m\in\mcA$. 
For such $S$, it is clear that
\begin{equation}\label{eq:disordered_k_m_lemma_1_2_eq1}
    \sigma^{-n}(S)\cap\pi_n^{-1}\{(a_1, \dots, a_n)\}
        = 
    \pi_{m+n}^{-1}\{(a_1, \dots, a_n, b_1, \dots, b_m)\}, 
\end{equation}
so for any $\vartheta\in\rstates$, \Cref{Lem:The quantum measure} gives
\begin{align}
    \mbQ^*_\omega\seq{\sigma^{-n}(S)\cap\pi_n^{-1}\{(a_1, \dots, a_n)\}}
        &=
    \mbQ^*_\omega\seq{\pi_{m+n}^{-1}\{(a_1, \dots, a_n, b_1, \dots, b_m)\}}
        \\
        &= 
    T_{a_1; \theta(\omega)}^\dagger\circ\cdots\circ T_{a_n; \theta^n(\omega)}^\dagger\circ
    T_{b_1; \theta^{n+1}(\omega)}\circ\cdots\circ T_{b_m; \theta^{m+n}(\omega)}(I)
        \\
        &= 
    T_{a_1; \theta(\omega)}^\dagger\circ\cdots\circ T_{a_n; \theta^n(\omega)}^\dagger
    \seq{
        \mbQ^*_{\theta^n(\omega)}\seq{\pi^{-1}_m\{(b_1,\ldots b_m)\}}
    }
        \\
        &= 
    T_{a_1; \theta(\omega)}^\dagger\circ\cdots\circ T_{a_n; \theta^n(\omega)}^\dagger
    \seq{
        \mbQ^*_{\theta^n(\omega)}\seq{S}
    },
\end{align}
which is what we wanted. 
\end{proof}
As a corollary, we have the following. 
Recall that 
\begin{equation}
    \Phi^{(n)}_\omega = \phi_{\theta^n(\omega)}\circ\cdots\circ\phi_{\theta(\omega)} \quad \text{and} \quad \Phi^{(-n)}_\omega = \phi_\omega\circ\cdots\circ\phi_{\theta^{1-n}(\omega)} 
\end{equation}
for $n \in \mbN$.
\begin{lemma}\label{lem:Sigma_shift_property_of_quantum_measure}
 Let $E\in\Sigma$ and $n\in\mbN$. 
Then
 \begin{equation}\label{eq:lem:Sigma_shift_property_of_quantum_measure_2}
                    \mbQ_{\omega}^*\seq{\sigma^{-n}(E)}
                        =
                            \Phi^{(n)\dagger}_{\omega}\seq{\mbQ^*_{\theta^n(\omega)}(E)} 
\end{equation}
for $\pr$-almost every $\omega\in\Omega$. 
Consequently, for any $\vartheta\in\rstates$, we have that 
\begin{equation}\label{eq:lem:Sigma_shift_property_of_quantum_measure_1}
    \mbQ_{\vartheta; \omega}\seq{\sigma^{-n}(E)}
        =
    \mbQ_{\Phi^{(-n)}(\vartheta_{-n}); \theta^n(\omega)}\seq{E}
\end{equation}
for $\pr$-almost every $\omega\in\Omega$, where $\vartheta_{-n}\in\rstates$ is defined by $\vartheta_{-n; \omega} = \vartheta_{\theta^{-n}(\omega)}$.
\end{lemma}
\begin{proof}
\Cref{eq:lem:Sigma_shift_property_of_quantum_measure_1} follows immediately from \Cref{eq:lem:Sigma_shift_property_of_quantum_measure_2} by the computation 
\begin{align}
     \mbQ_{\vartheta; \omega}\seq{\sigma^{-n}(E)}
        =
     \inner{\vartheta_\omega}{\mbQ^*_\omega\seq{\sigma^{-n}(E)}}
        &= 
     \inner{\vartheta_\omega}{\Phi^{(n)\dagger}_{\omega}\seq{\mbQ^*_{\theta^n(\omega)}(E)}}\\
        &= 
    \inner{\Phi^{(n)}_{\omega}\seq{\vartheta_\omega}}{{\mbQ^*_{\theta^n(\omega)}(E)}}
        =
     \mbQ_{\Phi^{(-n)}(\vartheta_{-n}); \theta^n(\omega)}\seq{E},
\end{align}
so it suffices to prove \Cref{eq:lem:Sigma_shift_property_of_quantum_measure_2}. 
For $E\in\Sigma$, we can write express $\sigma^{-n}(E)$ as the disjoint union
\begin{equation}
    \sigma^{-n}(E)
        =
    \bigsqcup_{(a_1, \dots, a_n)\in\mcA^n}
    \sigma^{-n}(E)\cap \pi_n^{-1}\{(a_1, \dots, a_n)\}.
\end{equation}
So, by the previous lemma and the additivity of the matrix-valued measure $\mbQ^*$, we conclude that 
\begin{align}
    \mbQ^*_\omega\seq{\sigma^{-n}(E)}
        &=
    \sum_{(a_1, \dots, a_n)\in\mcA^n}
     T_{a_1; \theta(\omega)}^\dagger\circ\cdots\circ T_{a_n; \theta^n(\omega)}^\dagger
    \seq{
        \mbQ^*_{\theta^n(\omega)}\seq{E}
    }
        \\
        &= 
    \Phi^{(n)\dagger}_\omega\seq{\mbQ^*_{\theta^n(\omega)}\seq{E}}
\end{align}
by the definition of $\Phi^{(n)}_\omega$.
\end{proof}
Recall that we say $\rho\in\rstates$ is stochastically stationary under $\mcV$ if 
\begin{equation}
    \phi_{\theta(\omega)}\seq{\rho_\omega}
    =
    \rho_{\theta(\omega)}
\end{equation}  
holds $\pr$-almost surely, where $\phi = \phi_\mcV$.
The previous lemmas allow us to see this condition as sufficient to ensure that the annealed quantum measure $\bQ_\rho$ is $\tau$-invariant. 
\begin{lemma}\label{lemma:tau_preserves_annealed}
    Let $\mcV$ be a random Kraus ensemble and let $\rho\in\rstates$ be stochastically stationary under $\phi_\mcV$. 
    Then $\tau: \Omega\times\outcomes\to\Omega\times\outcomes$ is a $\bQ_\rho$-probability preserving transformation.
\end{lemma}
\begin{proof}
    Let $\Gamma\in\mcF\otimes\Sigma$. 
    Note that, for any $\omega\in\Omega$ and $n\in\mbN$, we have 
    \begin{align}
        \seq{\tau^{-n}(\Gamma)}^\omega
            &=
        \left\{
            \bar{a}\in\mcA^\mbN
                \,\,:\,\,
            \sigma^n(\bar{a})\in\Gamma^{\theta^n(\omega)}
        \right\}\\
            &= 
        \sigma^{-n}\seq{\Gamma^{\theta^n(\omega)}}\label{Eq:Tau shift and sigma shift}.
    \end{align}
    Now, because $\theta$ is $\pr$-preserving and $\rho$ is stochastically stationary under $\phi$, we find that 
    \begin{align}
        \bQ_{\rho}(\Gamma)
            =
        \int_\Omega 
            \mbQ_{\rho; \omega}\seq{\Gamma^\omega}
            \,
            \pr(\dee\omega)
            &= 
        \int_\Omega 
            \mbQ_{\rho; \theta(\omega)}\seq{\Gamma^{\theta(\omega)}}
            \,
            \pr(\dee\omega)\\
            &= 
        \int_\Omega 
            \inner{\rho_{\theta(\omega)}}{\mbQ^*_{\theta(\omega)}\seq{\Gamma^{\theta(\omega)}}}
            \,
            \pr(\dee\omega)\\
            &= 
        \int_\Omega 
            \inner{
            \phi_{\theta(\omega)}(\rho_\omega)
            }{\mbQ^*_{\theta(\omega)}\seq{\Gamma^{\theta(\omega)}}}
            \,
            \pr(\dee\omega)\\
            &= 
        \int_\Omega 
            \mbQ_{\phi(\rho_{-1});\theta(\omega)}\seq{\Gamma^{\theta(\omega)}}
            \,
            \pr(\dee\omega),
    \end{align}
    where $\rho_{-1; \omega} = \rho_{\theta^{-1}(\omega)}$.
    From \Cref{lem:Sigma_shift_property_of_quantum_measure} and \Cref{Eq:Tau shift and sigma shift}, we conclude 
    \begin{align}
        \bQ_{\rho}(\Gamma)
            = 
        \int_\Omega 
           \mbQ_{\rho; \omega}\seq{\sigma^{-1}\seq{\Gamma^{\theta(\omega)}}}
           \,
           \pr(\dee\omega)
           &= 
        \int_\Omega 
           \mbQ_{\rho; \omega}\seq{\seq{\tau^{-1}\seq{\Gamma}}^\omega}
           \,
           \pr(\dee\omega)\\
        &= 
        \bQ_\rho\seq{\tau^{-1}\seq{\Gamma}},
    \end{align}
    which is what we wanted to show. 
\end{proof}
        In fact, the ergodicity of $\theta$ for $\pr$ allows us to conclude the following. 
        \begin{lemma}\label{lemma:qunched_a.s._constant}
        Let $\mcV$ be a random Kraus ensemble and let $\rho\in\rstates$ be stochastically stationary under $\phi_\mcV$. 
        Then for any $\tau$-invariant $\Gamma\in\mcF\otimes\Sigma$, we have that
        \begin{equation}
            \mbQ_{\rho; \omega}\seq{\Gamma^\omega}
                =
            \bQ_{\rho}\seq{\Gamma}
        \end{equation}
        for $\pr$-almost every $\omega\in\Omega$. 
        \end{lemma}
        \begin{proof}
            Arguing as in the previous lemma, we use the fact that $\rho$ is stochastically stationary under $\phi$ together with \Cref{lem:Sigma_shift_property_of_quantum_measure} and \Cref{Eq:Tau shift and sigma shift} to see that $\mbQ_{\rho; \theta(\omega)}\seq{\Gamma^{\theta(\omega)}}
            =
            \mbQ_{\rho; \omega}\seq{\seq{\tau^{-1}\seq{\Gamma}}^\omega}$ for $\pr$-almost every $\omega\in\Omega$. 
            %
            % \begin{align}
            %     \mbQ_{\rho; \theta(\omega)}\seq{\Gamma^{\theta(\omega)}}
            %         = 
            %     \inner{\rho_{\theta(\omega)}}{\mbQ^*_{\theta(\omega)}\seq{\Gamma^{\theta(\omega)}}}
            %         &= 
            %     \inner{\rho_\omega}{
            %     \phi_{\mcV; \theta(\omega)}^\dagger\seq{
            %     \mbQ^*_{\theta(\omega)}\seq{\Gamma^{\theta(\omega)}}}
            %     }\\
            %         &= 
            %     \inner{\rho_\omega}{
            %     \mbQ^*_\omega\seq{
            %     \sigma^{-1}
            %     \seq{\Gamma^{\theta(\omega)}}}
            %     }\\
            %         &= 
            %     \inner{\rho_\omega}{
            %     \mbQ^*_\omega\seq{
            %     \seq{\tau^{-1}\seq{\Gamma}}^\omega}
            %     }.
            % \end{align}
            %
            Because $\Gamma$ is $\tau$-invariant, this gives 
            \begin{equation}
                 \mbQ_{\rho; \theta(\omega)}\seq{\Gamma^{\theta(\omega)}}
                =
                \mbQ_{\rho; \omega}\seq{\Gamma^\omega}. 
            \end{equation}
            Thus, the map $\omega\mapsto \mbQ_{\rho; \omega}\seq{\Gamma^\omega}$ is $\theta$-invariant, which by the $\pr$-ergodicity of $\theta$ allows us to conclude that
            \begin{equation}
                \mbQ_{\rho; \omega}\seq{\Gamma^\omega}
                =
                \int_\Omega
                \mbQ_{\rho; \omega}\seq{\Gamma^\omega}
                \,
                \pr(\dee\omega)
                =
                \bQ_{\rho}\seq{\Gamma}
            \end{equation}
            for $\pr$-almost every $\omega\in\Omega$, which concludes the proof. 
        \end{proof}
To prove the next lemma, we recall a result proved in \cite{ekblad2024ergodic}.
\begin{thmx}[\cite{ekblad2024ergodic}]\label{thm:dynamic_ergodicity}
            Given an ergodic and invertible system $(\Omega, \mcF, \pr, \theta)$ and a random quantum channel $\phi:\Omega\to\mapspace$ that is dynamically ergodic with unique stochastically stationary state $\SteadyStateNoOmega$, for any random state $\rho\in\rstates$ and random, essentially bounded observable $\mcO$ (i.e., $\mcO = \mcO^*$ and $\mcO\in L^\infty(\Omega, \matrices)$), we have
                \begin{equation}\label{eq:owen_thm_1}
                    \lim_{N\to\infty} 
                    \frac{1}{N} \sum_{n=1}^N
                        \inner{\phi_{\theta^n(\omega)}\circ\ldots\circ \phi_{\theta(\omega)}(\rho_\omega)}{\mcO_{\theta^n(\omega)}} 
                                    =
                    \int_\Omega 
                    \inner{\SteadyState{\omega'}}{
                    \mcO_{\omega'}
                    }
                    \,
                    \pr(\dee\omega').
                \end{equation}
        \end{thmx}
We may now prove our last technical lemma. 
 \begin{lemma}\label{lemma:qunched_is_constant_for_any}
Let $\mcV$ be a dynamically ergodic random Kraus ensemble and let $\SteadyStateNoOmega\in\rstates$ denote its unique stochastically stationary state.  
Then for any $\Gamma\in\mcF\otimes\Sigma$, $\tau^{-1}\seq{\Gamma} = \Gamma$ implies that 
\begin{equation}\label{eqn:qunched_is_constant_for_any_eqn1}
    \mbQ^*_\omega\seq{\Gamma^\omega}
        =
    \bQ_{\SteadyStateNoOmega}(\Gamma)\mbI 
\end{equation}
for $\pr$-almost every $\omega\in\Omega$. 
In particular, for any $\vartheta\in\rstates$, 
\begin{equation}\label{eqn:qunched_is_constant_for_any_eqn2}
    \mbQ_{\vartheta; \omega}\seq{\Gamma}
        =
    \bQ_{\SteadyStateNoOmega}(\Gamma)
\end{equation}
for $\pr$-almost every $\omega\in\Omega$. 
\end{lemma}
\begin{proof}
\Cref{eqn:qunched_is_constant_for_any_eqn2} follows immediately from \Cref{eqn:qunched_is_constant_for_any_eqn1}, so we prove \Cref{eqn:qunched_is_constant_for_any_eqn1}.
    Since $\Gamma$ is $\tau$-invariant, we have 
    \begin{equation}
        \Gamma^\omega 
        =
        \seq{\tau^{-n}\seq{\Gamma}}^\omega
        =
        \sigma^{-n}\seq{\Gamma^{\theta^n(\omega)}}
    \end{equation}
    for any $\omega\in\Omega$. 
    So, for any (nonrandom) $\psi\in\states$, we have that 
    \begin{align}    
        \inner{\psi}{\mbQ_\omega^*\seq{\Gamma^\omega}}
            &=
       \frac{1}{N}
       \sum_{n=1}^N
        \inner{\psi}{\mbQ_\omega^*\seq{\sigma^{-n}\seq{\Gamma^{\theta^n(\omega)}}}}
                \\
            &= 
        \frac{1}{N}
       \sum_{n=1}^N
        \inner{\Phi^{(n)}_\omega(\psi)}{
        \mbQ^*_{\theta^n(\omega)}\seq{\Gamma^{\theta^n(\omega)}}
        }
    \end{align}
    by \Cref{lem:Sigma_shift_property_of_quantum_measure}.
    In particular, noting that the random matrix defined by $\mcO_{\omega'}:= \mbQ^*_{\omega'}(\Gamma^{\omega'})$ is positive-semidefinite almost surely and satisfies $\|\mcO_{\omega'}\|\leq 1$, we may apply \Cref{thm:dynamic_ergodicity} to conclude from the above that for almost every $\omega\in\Omega,$ we have 
    \begin{align}
         \inner{\psi}{\mbQ_\omega^*\seq{\Gamma^\omega}}
         &=
         \lim_{N\to\infty}
         \frac{1}{N}
       \sum_{n=1}^N
        \inner{\Phi^{(n)}_\omega(\psi)}{
        \mbQ^*_{\theta^n(\omega)}\seq{\Gamma^{\theta^n(\omega)}}
        }
        \\
        &=
        \int_\Omega 
        \inner{\SteadyState{\omega'}}{\mbQ^*_{\omega'}\seq{\Gamma^{\omega'}}}\,
        \pr(\dee\omega')\\
        &= 
        \bQ_{\SteadyStateNoOmega}\seq{\Gamma}.
    \end{align}
    Because $\psi\in\states$ was arbitrary, we conclude from this that the matrix $\mbQ^*_\omega\seq{\Gamma^\omega}$ is equal to $\bQ_{\SteadyStateNoOmega}\seq{\Gamma}\mbI$, concluding the proof. 
\end{proof}
    \subsection{Proofs of theorems}
    We are now equipped to prove our theorems. 
    Before we proceed further, we recall the following theorem due to Beck \& Schwartz \cite{beck1957vector}. 
        \begin{thmx}[\cite{beck1957vector}]\label{Thm:Beck and Schwartz}
            Let $\scrX$ be a reflexive Banach space and let $(S, \mcG, m)$ be a $\sigma$-finite measure space. 
            Let $s\mapsto \psi_s$ be a strongly measurable function taking values in $\scrB(\scrX)$. 
            Suppose that $\|\psi_s\|\leq 1$ for all $s\in S$. 
            Let $h$ be a measure-preserving transformation in $(S, \mcG, m)$. 
            Then for each $X\in L^1(S, \scrX)$, there is $\bar{X}\in L^1(S, \scrX)$ such that 
                \begin{equation}
                    \lim_{M\to\infty}
                    \frac{1}{M}
                    \sum_{N=1}^M
                    \psi_s\cdots \psi_{h^{N-1}(s)}
                        \big(X_{h^N(s)}\big)
                        =
                        \bar{X}_s
                \end{equation}
            strongly in $\scrX$ almost everywhere in $S$, and $\bar{X}_s = \psi_s\big(\bar{X}_{h(s)}\big)$ almost everywhere in $S$. 
            Moreover, if $m(S) < \infty$, then $\bar{X}$ is also the limit in the mean of order 1. 
        \end{thmx}
        We begin with the proof of \Cref{Thm:lln_for_Q}, restated here for convenience. 
\thmlln*
\begin{proof}
    By \Cref{Thm:Beck and Schwartz} applied to the map $\omega\mapsto \phi_\omega$ and $h = \theta^{-1}$, we see that for any $\vartheta\in\rstates$, the $\pr$-almost sure limit 
    \begin{equation}
    \overline{\vartheta}_\omega 
        :=
        \lim_{N\to\infty}
        \frac{1}{N}
        \sum_{n=1}^N
       \Phi^{(-n)}_\omega\seq{\vartheta_{\theta^{-n}(\omega)}}
        \label{Eqn:LLN Annealed, Eqn 1}
    \end{equation}
    exists and is a stochastically stationary quantum state for $\mcV$. 
    By the assumption of dynamical ergodicity, we have that $\overline{\vartheta}_\omega = \SteadyState{\omega}$ for $\pr$-almost every $\omega\in\Omega$. 
    We now compute 
    \begin{align}
        \bQ_\vartheta\seq{\tau^{-n}\seq{\Gamma}}
            &=
        \int_\Omega 
        \mbQ_{\vartheta_\omega}\seq{\seq{\tau^{-n}\seq{\Gamma}}^\omega}
        \,
        \pr(\dee\omega)\\
            &=
        \int_\Omega 
         \mbQ_{\vartheta_\omega}\seq{
         \sigma^{-n}\seq{\Gamma^{\theta^n(\omega)}}
         }
        \,
        \pr(\dee\omega)
            &&\text{by \Cref{Eq:Tau shift and sigma shift}}\\
            &= 
        \int_\Omega 
        \inner{\vartheta_\omega}
        {
        \Phi^{(n)\dagger}_\omega\seq{\mbQ^*_{\theta^n(\omega)}\seq{\Gamma^{\theta^n(\omega)}}}
        }
        \,
        \pr(\dee\omega)
            &&\text{by \Cref{lem:Sigma_shift_property_of_quantum_measure}}\\
            &= 
        \int_\Omega 
        \inner{
        \Phi^{(n)}_\omega\seq{\vartheta_\omega}}
        {
        \mbQ^*_{\theta^n(\omega)}\seq{\Gamma^{\theta^n(\omega)}}
        }
        \,
        \pr(\dee\omega)\\
            &= 
        \int_\Omega 
        \inner{
       \Phi^{(-n)}_\omega\seq{\vartheta_{\theta^{-n}(\omega)}}
       }
        {
        \mbQ^*_{\omega}\seq{\Gamma^{\omega}}
        }
        \,
        \pr(\dee\omega),
        \label{Eqn:LLN Annealed, Eqn 2}
    \end{align}
    where the last equality holds by the change of variables $\omega\mapsto\theta^{-n}(\omega)$ and uses the fact that $\theta^{-1}$ is $\pr$-preserving. 
    In conjunction with the fact that $\overline{\vartheta}_\omega = \SteadyState{\omega}$ almost surely, if we apply the tracial H\"older's inequality to see that
    \begin{equation}
        \left|\inner{
        \frac{1}{N}
        \sum_{n=1}^N
        \Phi^{(-n)}_\omega\seq{\vartheta_{\theta^{-n}(\omega)}}
        }
        {
        \mbQ^*_{\omega}\seq{\Gamma^{\omega}}
        }\right|
        \leq 
        1
    \end{equation}
    holds for $\pr$-almost every $\omega\in\Omega$, 
    then we may apply the dominated convergence theorem to conclude that
    \begin{align}
        \lim_{N\to\infty}
        \frac{1}{N}
        \sum_{n=1}^N
        \bQ_\vartheta\seq{\tau^{-n}\seq{\Gamma}}
            &=
        \lim_{N\to\infty}
        \frac{1}{N}
        \sum_{n=1}^N
        \int_\Omega 
        \inner{
       \Phi^{(-n)}_\omega\seq{\vartheta_{\theta^{-n}(\omega)}}}
        {
        \mbQ^*_{\omega}\seq{\Gamma^{\omega}}
        }
        \,
        \pr(\dee\omega)\\
            &=
        \int_\Omega 
        \inner{
         \lim_{N\to\infty}
        \frac{1}{N}
        \sum_{n=1}^N
       \Phi^{(-n)}_\omega\seq{\vartheta_{\theta^{-n}(\omega)}}}
        {
        \mbQ^*_{\omega}\seq{\Gamma^{\omega}}
        }
        \,
        \pr(\dee\omega)\\
            &=
        \int_\Omega 
        \inner{
         \SteadyState{\omega}
         }
        {
        \mbQ^*_{\omega}\seq{\Gamma^{\omega}}
        }
        \,
        \pr(\dee\omega)\\
            &=
       \bQ_{\SteadyStateNoOmega}\seq{\Gamma},
    \end{align}
    as desired. 
\end{proof}
We now prove \texorpdfstring{\Cref{Thm:Annealed_Erg_Thm_1}}{l}. 
Recall its statement. 
\thmAnnnealedErgodicity*
\begin{proof}
We have already shown that $\bQ_{\SteadyStateNoOmega}\seq{\tau^{-1}\seq{\Gamma}} = \bQ_{\SteadyStateNoOmega}\seq{\Gamma}$ in \Cref{lemma:tau_preserves_annealed}, and from \Cref{lemma:qunched_is_constant_for_any} we see that, for any $\tau$-invariant $\Gamma\in\mcF\otimes\Sigma$ and any $\vartheta\in\rstates$, we have that 
\begin{align}
    \bQ_{\vartheta}\seq{\Gamma}
    =
    \int_\Omega 
    \mbQ_{\vartheta; \omega}\seq{\Gamma^\omega}
    \,\pr(\dee\omega)
    &= 
    \bQ_{\SteadyStateNoOmega}(\Gamma),
\end{align}
so all that remains is to show the claimed ergodic property of $\tau$ with respect to $\bQ_{\SteadyStateNoOmega}$.
We begin by showing that for any $\tau$-invariant $\Gamma\in\mcF\otimes\Sigma$, the equality
\begin{equation}\label{Eqn:Annealed erg thm intersection}
    \bQ_{\vartheta}\seq{\Gamma\cap\Gamma'}
    =
    \bQ_{\SteadyStateNoOmega}(\Gamma)\cdot\bQ_{\vartheta}\seq{\Gamma'}
\end{equation}
holds for any $\vartheta\in\rstates$. 
To see this, it suffices by a monotone class argument to show that \Cref{Eqn:Annealed erg thm intersection} holds for $\Gamma'$ of the form $\Gamma' = F\times E$ with $F\in\mcF$ and $E = \pi_{n}^{-1}\{(a_1, \dots, a_n)\}$ for all $n\in\mbN$ and $a_1, \dots, a_n\in\mcA$. 
Noting that $\seq{F\times E}^\omega = E$ and $\seq{\Gamma_1\cap\Gamma_2}^\omega = \Gamma_1^\omega\cap\Gamma_2^\omega$ for any $\Gamma_1, \Gamma_2\in\mcF\otimes\Sigma$, we find that 
\begin{align}
    \bQ_{\vartheta}\seq{
    \Gamma\cap\seq{F\times E}
    }
        &=
    \bQ_{\vartheta}\seq{
    \tau^{-n}\seq{\Gamma}\cap\seq{F\times E}
    }
            \\
        &= 
    \int_\Omega 
    \mbQ_{\vartheta; \omega}\seq{
    \seq{
        \tau^{-n}\seq{\Gamma}\cap\seq{F\times E}
    }
    ^\omega
    }
    \,
    \pr(\dee\omega)
            \\
        &= 
    \int_\Omega 
    \mbQ_{\vartheta; \omega}\seq{
        \seq{\tau^{-n}\seq{\Gamma}}^\omega\cap E
    }
    \,
    \pr(\dee\omega)
            \\
        &= 
    \int_\Omega 
    \mbQ_{\vartheta; \omega}\seq{
        \seq{\sigma^{-n}\seq{\Gamma^{\theta^n(\omega)}}}\cap E
    }
    \,
    \pr(\dee\omega)
\end{align}
So, since $E = \pi_n^{-1}\{(a_1, \dots, a_n)\}$, \Cref{lem:disordered_k_m_lemma_1_2} gives
\begin{align}
     \bQ_{\vartheta}\seq{
    \Gamma\cap\seq{F\times E}
    }
        &= 
    \int_\Omega 
    \inner{
    \vartheta_\omega
        }{
    T_{a_1; \theta(\omega)}^\dagger\circ\cdots\circ T_{a_n; \theta^n(\omega)}^\dagger
        \seq{
        \mbQ^*_{\theta^n(\omega)}\seq{\Gamma^{\theta^n(\omega)}}
        }
    }
    \,
    \pr(\dee\omega)
            \\
        &= 
    \int_\Omega 
    \inner{
    T_{a_n; \theta^n(\omega)}\circ\cdots\circ T_{a_1; \theta(\omega)}
    \seq{\vartheta_\omega}
        }{
        \mbQ^*_{\theta^n(\omega)}\seq{\Gamma^{\theta^n(\omega)}}
    }
    \,
    \pr(\dee\omega).
\end{align}
Thus, by \Cref{lemma:qunched_is_constant_for_any}, we conclude from the fact that $\Gamma$ is $\tau$-invariant that
\begin{align}
    \bQ_{\vartheta}\seq{
    \Gamma\cap\seq{F\times E}
    }
        &=
        \bQ_{\SteadyStateNoOmega}(\Gamma)
     \int_\Omega 
    \inner{
    T_{a_n; \theta^n(\omega)}\circ\cdots\circ T_{a_1; \theta(\omega)}
    \seq{\vartheta_\omega}
        }{
        I
    }
    \,
    \pr(\dee\omega)
            \\
        &= 
     \bQ_{\SteadyStateNoOmega}(\Gamma)
     \int_\Omega
     \mbQ_{\vartheta; \omega}\seq{\pi_n^{-1}\{(a_1, \dots, a_n)\}}
    \,
    \pr(\dee\omega)
            \\
        &= 
     \bQ_{\SteadyStateNoOmega}(\Gamma)
     \bQ_{\vartheta}\seq{F\times E}, 
\end{align}
which is what we wanted.
Thus, the claim is indeed true. 
It is now immediate that the theorem follows from the claim: indeed, by letting $\vartheta = \SteadyStateNoOmega$ and $\Gamma' = \Gamma$ in \Cref{Eqn:Annealed erg thm intersection}, we find that, for any $\tau$-invariant $\Gamma$, 
\begin{equation}
    \bQ_{\SteadyStateNoOmega}(\Gamma) = \bQ_{\SteadyStateNoOmega}(\Gamma)^2, 
\end{equation}
which establishes that $\bQ_{\SteadyStateNoOmega}(\Gamma)\in\{0, 1\}$, and concludes the proof. 
\end{proof}
With this theorem in hand, the proof of \Cref{Thm:Quenched_Ergodic} is straightforward. 
Recall its statement. 
\thmQuenchedErg*
\begin{proof}
    Let $E\in\Sigma$. 
    Then from \Cref{Thm:Annealed_Erg_Thm_1}, we have that 
    \begin{align}
        \avg{\pr}{\mbQ_{\SteadyStateNoOmega}}\seq{\sigma^{-1}(E)}
            =
        \bQ_{\SteadyStateNoOmega}\seq{\tau^{-1}\seq{\Omega\times E}}
        =
        \bQ_{\SteadyStateNoOmega}\seq{\Omega\times E}
        =
        \avg{\pr}{\mbQ_{\SteadyStateNoOmega}}\seq{E}.
    \end{align}
    Moreover, if $\sigma^{-1}(E) = E$, then $\tau^{-1}\seq{\Omega\times E} = \Omega\times E$, so by \Cref{Thm:Annealed_Erg_Thm_1} it holds that 
    \begin{equation}
        \{0, 1\}\ni \bQ_{\SteadyStateNoOmega}\seq{\Omega\times E}
        =
        \avg{\pr}{\mbQ_{\SteadyStateNoOmega}}\seq{E}
    \end{equation}
    for any such $E$. 
    Lastly, by \Cref{lemma:qunched_is_constant_for_any}, if $\sigma^{-1}(E) = E$ hence $\tau^{-1}(\Omega\times E) = \Omega\times E$, we see that 
    \begin{align}
        \mbQ_{\vartheta; \omega}\seq{E}
        =
        \mbQ_{\vartheta; \omega}\seq{\seq{\Omega\times E}^\omega}
        =
        \bQ_{\SteadyStateNoOmega}\seq{\Omega\times E}
        =
        \avg{\pr}{\mbQ_{\SteadyStateNoOmega}}\seq{E},
    \end{align}
    which concludes the proof. 
\end{proof}
Finally, we prove \Cref{Thm:Path averages equal quantum averages of equilibrium state}. 
First, we recall the relevant notation. 
Given $(b_1, \dots, b_m)\in\mcA^m$ and $n\in\mbN$, let $A_{n}(b_1, \dots, b_m)$ denote
\begin{equation}
    A_{n}(b_1, \dots, b_m)
        :=
    \Big\{
        \seq{a_k}_{k\in\mbN}\in\mcA^\mbN
            \,\,:\,\,
        a_{n} = b_1, \dots, a_{n+m-1} = b_m
    \Big\},
\end{equation}
the set of sequences of measurement outcomes in which $(b_1, \dots, b_m)$ occurs starting at time $n$.
\thmcoutinglago*
\begin{proof}
Fix $m\in\mbN$ and $b_1, \dots, b_m\in\mcA$. 
Let $f:\mcA^\mbN\to\{0, 1\}$ denote the $\Sigma$-measurable function 
\begin{equation}
    f\seq{\bar{a}}
        :=
    1_{\{(b_1, \dots, b_m)\}}\seq{\pi_m(\bar{a})}. 
\end{equation}
Then notice that
\begin{equation}
    \dfrac{\#\Big\{n < N \,\,:\,\, \bar{a} \in A_n(b_1, \dots, b_m)\Big\}}{N} = \frac{1}{N} \sum_{n=1}^N f(\sigma^n(\bar{a}))
\end{equation}
for any $N\in\mbN$. 
Thus, by applying Birkhoff's pointwise ergodic theorem, we may conclude from \Cref{Thm:Quenched_Ergodic} that 
\begin{align}
    \lim_{N\to\infty}
    \dfrac{\#\Big\{n < N \,\,:\,\, \bar{a} \in A_n(b_1, \dots, b_m)\Big\}}{N}   &=
    \lim_{N\to\infty}
    \frac{1}{N} \sum_{n=1}^N f(\sigma^n(\bar{a}))
            \\
        &= 
    \int_{\mcA^\mbN}
    f(\bar{b})\,\,
    \mbE_{\pr}[\mbQ_{\SteadyStateNoOmega}](\dee\bar{b})
            \\
        &=
    \mbE_{\pr}[\mbQ_{\SteadyStateNoOmega}](b_1, \dots, b_m)
\end{align}
holds for every $\bar{a}\in E$, where $E\in\Sigma$ is a set satisfying $\sigma^{-1}\seq{E} = E$ and $\mbE_{\pr}[\mbQ_{\SteadyStateNoOmega}](E) = 1$. 
In particular, because $\sigma^{-1}\seq{E} = E$, another application of \Cref{Thm:Quenched_Ergodic} yields that 
\begin{equation}
    \mbQ_{\vartheta; \omega}(E)
        =
    \mbE_{\pr}[\mbQ_{\SteadyStateNoOmega}](E)
        =
    1
\end{equation}
for every $\vartheta\in\rstates$ and $\pr$-almost every $\omega\in\Omega$, which concludes the proof. 
\end{proof}